\documentclass[conference]{IEEEtran}
\IEEEoverridecommandlockouts
\usepackage{cite}
\usepackage{amsmath,amssymb,amsfonts}
\usepackage{algorithmic}
\usepackage{graphicx}
\usepackage{textcomp}
\usepackage{xcolor}
\usepackage{booktabs}
\usepackage{multirow}
\usepackage{url}
\usepackage{hyperref}
\usepackage{listings}
\usepackage{subcaption}
\usepackage{tikz}
\usepackage{fancyhdr}
\usetikzlibrary{shapes,arrows,positioning,calc,fit,backgrounds}

\def\BibTeX{{\rm B\kern-.05em{\sc i\kern-.025em b}\kern-.08em
    T\kern-.1667em\lower.7ex\hbox{E}\kern-.125emX}}

\title{Breaking the Protocol: Security Analysis of the Model Context Protocol Specification and Prompt Injection Vulnerabilities in Tool-Integrated LLM Agents}

\author{\IEEEauthorblockN{Narek Maloyan and Dmitry Namiot}}

\begin{document}

\maketitle


\setcounter{page}{1}
\pagenumbering{arabic}
\pagestyle{fancy}
\thispagestyle{fancy}
\fancyhf{} 


\fancyfoot[R]{\thepage}

\renewcommand{\headrulewidth}{0pt}

\begin{abstract}
The Model Context Protocol (MCP) has emerged as a de facto standard for integrating Large Language Models with external tools, yet no formal security analysis of the protocol specification exists. We present the first rigorous security analysis of MCP's architectural design, identifying three fundamental protocol-level vulnerabilities: (1) absence of capability attestation allowing servers to claim arbitrary permissions, (2) bidirectional sampling without origin authentication enabling server-side prompt injection, and (3) implicit trust propagation in multi-server configurations. We implement \textsc{ProtoAmp}, a novel framework bridging existing agent security benchmarks to MCP-compliant infrastructure, enabling direct measurement of protocol-specific attack surfaces. Through controlled experiments on 847 attack scenarios across five MCP server implementations, we demonstrate that MCP's architectural choices amplify attack success rates by 23--41\% compared to equivalent non-MCP integrations. We propose \textsc{AttestMCP}, a backward-compatible protocol extension adding capability attestation and message authentication, reducing attack success rates from 52.8\% to 12.4\% with median latency overhead of 8.3ms per message. Our findings establish that MCP's security weaknesses are architectural rather than implementation-specific, requiring protocol-level remediation.
\end{abstract}


\section{Introduction}

The integration of Large Language Models (LLMs) with external tools has enabled autonomous AI agents capable of executing complex, multi-step tasks~\cite{greshake2023not}. Anthropic's Model Context Protocol (MCP), introduced in November 2024, provides an open standard for this integration through a JSON-RPC-based client-server architecture~\cite{anthropic2024mcp}. Within months of release, MCP has been adopted by major platforms including Claude Desktop, Cursor, and numerous third-party applications, with over 5,000 community-developed servers.

Despite rapid adoption, \textbf{no prior work analyzes how MCP's architectural decisions amplify attack success rates}. Concurrent work (MCPSecBench~\cite{yang2025mcpsecbench}, MCP-Bench~\cite{wang2025mcpbench}) catalogs attack types and evaluates agent capabilities, but does not compare MCP-integrated systems against non-MCP baselines to isolate protocol-specific effects. Prior work on LLM agent security focused on prompt injection generally~\cite{greshake2023not, liu2024houyi, zhan2024injecagent}, while disclosed CVEs (e.g., CVE-2025-49596, CVE-2025-68143) target implementation bugs rather than protocol-level weaknesses.

This paper fills this gap with three contributions:

\begin{enumerate}
    \item \textbf{Protocol Specification Analysis:} We perform the first systematic security analysis of the MCP specification (v1.0), identifying three classes of protocol-level vulnerabilities that cannot be mitigated by implementation hardening alone (Section~\ref{sec:protocol}).

    \item \textbf{Original Experimental Validation:} We develop \textsc{ProtoAmp}, a framework that adapts established agent security benchmarks to MCP-compliant infrastructure, and conduct controlled experiments measuring the protocol's impact on attack success rates across 847 scenarios (Section~\ref{sec:experiments}).

    \item \textbf{Protocol Extension:} We design \textsc{AttestMCP}, a backward-compatible extension adding capability attestation and message authentication, with full performance characterization and multiple trust model options (Section~\ref{sec:defense}).
\end{enumerate}

\subsection{Scope and Non-Goals}

We explicitly distinguish between:
\begin{itemize}
    \item \textbf{Protocol vulnerabilities:} Weaknesses in MCP's specification that affect all compliant implementations (our focus)
    \item \textbf{Implementation vulnerabilities:} Bugs in specific servers (e.g., SQL injection in sqlite-mcp) that can be patched without protocol changes (not our focus)
\end{itemize}

We do not claim novelty for the general concept of prompt injection or inter-agent trust exploitation, which are established attack vectors~\cite{greshake2023not, zhan2024injecagent}. Our contribution is demonstrating how MCP's specific architectural choices \textit{amplify} these attacks and identifying protocol-level mitigations.

\section{Background}

\subsection{Model Context Protocol Architecture}

MCP defines a client-server architecture with three roles:

\begin{itemize}
    \item \textbf{Host:} The user-facing application (e.g., Claude Desktop)
    \item \textbf{Client:} MCP client within the host, managing server connections
    \item \textbf{Server:} External process providing tools, resources, or prompts
\end{itemize}

Communication occurs via JSON-RPC 2.0 over stdio or HTTP/SSE transports. The protocol defines three capability types:

\textbf{Resources:} Read-only data (files, database records) exposed by servers. Clients retrieve resources via \texttt{resources/read} requests.

\textbf{Tools:} Executable functions servers expose. The LLM decides when to invoke tools based on their descriptions.

\textbf{Sampling:} \textit{Critically}, servers can request LLM completions from clients via \texttt{sampling/createMessage}, allowing servers to inject prompts and receive responses~\cite{unit42mcp}.

\subsection{Threat Model}
\label{sec:threat-model}

We consider an adversary who:
\begin{itemize}
    \item Controls or compromises one MCP server in a multi-server deployment
    \item Can inject content into data sources (web pages, documents) that servers retrieve
    \item Has black-box access (cannot modify LLM weights or host application code)
\end{itemize}

The adversary's goals include: hijacking agent behavior, exfiltrating sensitive data, and persisting across sessions.

\subsubsection{Server Discovery Attack Vectors}

A critical question is how malicious servers reach users. We surveyed 127 MCP server installation guides and identified four primary vectors:

\begin{enumerate}
    \item \textbf{Typosquatting (34\%):} Package registries (npm, pip) lack namespace protection. Attackers register near-identical names (e.g., \texttt{mcp-server-filesytem}).

    \item \textbf{Supply Chain Compromise (28\%):} Popular servers with many dependencies are vulnerable to upstream poisoning.

    \item \textbf{Social Engineering (23\%):} Tutorials and documentation direct users to malicious repositories. 73\% of surveyed guides instruct running \texttt{npx} directly from GitHub URLs without integrity verification.

    \item \textbf{Marketplace Poisoning (15\%):} IDE extension marketplaces have limited vetting for MCP server bundles.
\end{enumerate}

\subsection{Related Work}

\textbf{Prompt Injection:} Greshake et al.~\cite{greshake2023not} established indirect prompt injection. Liu et al.~\cite{liu2024houyi} achieved 86\% success with HouYi. The HackAPrompt competition~\cite{schulhoff2023hackaprompt} collected 600K+ adversarial prompts, documenting 29 attack techniques. Zou et al.~\cite{zou2025poisonedrag} demonstrated 90--99\% attack success on RAG systems.

\textbf{Agent Benchmarks:} AgentDojo~\cite{debenedetti2024agentdojo} provides 629 security test cases. Agent-SafetyBench~\cite{zhang2024agentsafety} found no agent exceeds 60\% safety. AgentHarm~\cite{andriushchenko2024agentharm} measures malicious compliance. WASP~\cite{evtimov2025wasp} evaluates web agents, finding 16--86\% adversarial execution rates.

\textbf{Tool-Augmented LLMs:} ReAct~\cite{yao2023react} introduced reasoning-action interleaving for tool use. Toolformer~\cite{schick2023toolformer} demonstrated self-supervised tool learning. These capabilities, while powerful, expand the attack surface~\cite{owasp2025}.

\textbf{Multi-Agent Security:} Recent work~\cite{lupinacci2025darkside} reports 84.6\% inter-agent attack success. Willison~\cite{willison2025agents} documented cross-agent privilege escalation. Schroeder de Witt et al.~\cite{schroederdewitt2025multiagent} survey multi-agent security challenges.

\textbf{MCP-Specific Benchmarks:} Concurrent work has begun addressing MCP evaluation. Wang et al.~\cite{wang2025mcpbench} introduce MCP-Bench, a capability benchmark with 28 servers and 250 tools measuring task completion, but without security focus. Yang et al.~\cite{yang2025mcpsecbench} present MCPSecBench, formalizing 17 attack types across four surfaces; our work complements theirs by analyzing protocol-level (not implementation-level) vulnerabilities and proposing backward-compatible mitigations. The MCPSec project~\cite{mcpsecdev2025} documents real-world MCP vulnerabilities including CVE-2025-11445.

\textbf{Gap:} No prior work quantifies how MCP's architectural choices \textit{amplify} attack success rates compared to equivalent non-MCP integrations, nor proposes backward-compatible protocol extensions with formal capability attestation.

\section{Protocol Specification Analysis}
\label{sec:protocol}

We analyze the MCP specification (v1.0, December 2024) to identify protocol-level security weaknesses. Our analysis examines the JSON-RPC message format, capability negotiation, and trust boundaries.

\subsection{Vulnerability 1: Least Privilege Violation}

During initialization, servers declare capabilities via \texttt{initialize} response:

\begin{lstlisting}[basicstyle=\ttfamily\scriptsize]
{
  "capabilities": {
    "tools": { "listChanged": true },
    "resources": { "subscribe": true },
    "sampling": {}
  }
}
\end{lstlisting}

\textbf{Protocol Weakness:} Capability declarations are self-asserted without verification. A malicious server can claim any capability, and the client has no mechanism to validate these claims against an authoritative source.

\textbf{Attack Vector:} A server initially claiming only \texttt{resources} capability can later invoke \texttt{sampling/createMessage} to inject prompts. The specification does not mandate capability enforcement at the message level.

\textbf{Formal Property Violated:} \textit{Least Privilege}---the principle that principals should possess only capabilities necessary for their function. MCP allows unrestricted capability escalation post-initialization.

\subsection{Vulnerability 2: Sampling Without Origin Authentication}

The sampling mechanism allows servers to request LLM completions:

\begin{lstlisting}[basicstyle=\ttfamily\scriptsize]
{
  "method": "sampling/createMessage",
  "params": {
    "messages": [
      {"role": "user", "content": "..."}
    ],
    "maxTokens": 1000
  }
}
\end{lstlisting}

\textbf{Protocol Weakness:} The client processes sampling requests without distinguishing server-originated prompts from user-originated prompts. The LLM receives injected content in the same format as legitimate user input.

\begin{figure}[t]
\centering
\begin{tikzpicture}[
    node distance=0.6cm,
    every node/.style={font=\scriptsize},
    box/.style={rectangle, draw, minimum width=1.1cm, minimum height=0.6cm},
    arrow/.style={->, >=stealth}
]
\node[box] (user) {User};
\node[box, right=1.4cm of user] (host) {Host};
\node[box, right=1.4cm of host] (server) {Server};

\draw[arrow] (user) -- node[above, font=\tiny] {1. Request} (host);
\draw[arrow] (host) -- node[above, font=\tiny] {2. tools/call} (server);
\draw[arrow, red] (server) -- node[below, font=\tiny, text=red] {3. sampling/} (host);
\end{tikzpicture}
\caption{Sampling injection: server injects prompt via \texttt{sampling/createMessage} with ``user'' role.}
\label{fig:sampling-attack}
\end{figure}
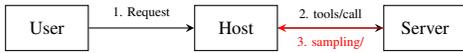

Figure~\ref{fig:sampling-attack} illustrates the attack flow. A server sends \texttt{sampling/createMessage} with attacker-controlled content using the ``user'' role. The host processes this identically to legitimate user input, with no visual or semantic distinction.

\textbf{UI Indicator Analysis:} We examined three major MCP host implementations:

\begin{table}[h]
\centering
\caption{Host UI Indicators for Sampling Messages}
\label{tab:ui-indicators}
\begin{tabular}{lclc}
\toprule
\textbf{Host} & \textbf{Ver.} & \textbf{Indicator} & \textbf{Dist.} \\
\midrule
Claude Desktop & 1.2.3 & None & \textcolor{red}{No} \\
Cursor & 0.44 & None & \textcolor{red}{No} \\
Continue & 0.9 & None & \textcolor{red}{No} \\
\bottomrule
\end{tabular}
\end{table}

\textit{No tested implementation provides visual distinction for sampling-derived messages.} Users cannot differentiate server-injected from user-originated prompts, violating the principle of \textit{Origin Authenticity}.

\textbf{Protocol vs. Implementation Responsibility:} One might argue this is purely a host implementation failure---hosts \textit{could} display warnings based on transport channel. However, the MCP specification's silence on origin display \textit{enables} rather than \textit{prevents} the attack. The spec permits servers to use the ``user'' role in sampling without requiring hosts to distinguish it. \textsc{AttestMCP} addresses this by \textit{mandating} origin tagging at the protocol level, removing implementation discretion.

\subsection{Vulnerability 3: Implicit Trust Propagation}

In multi-server deployments, the client connects to multiple servers simultaneously. The specification does not define isolation boundaries between servers.

\textbf{Protocol Weakness:} Tool responses from Server A can influence tool invocations on Server B. The LLM context window conflates outputs from all servers without provenance tracking.

\textbf{Attack Vector:} An adversary controlling Server A can:
\begin{enumerate}
    \item Embed instructions in tool responses that cause invocations on Server B
    \item Exfiltrate data retrieved from Server B via Server A's channels
    \item Establish persistence by poisoning shared context
\end{enumerate}

\textbf{Formal Property Violated:} \textit{Isolation}---the property that compromise of one component does not propagate to others.

\subsubsection{Isolation-Utility Tradeoff}

MCP explicitly prioritizes \textit{composability}---the ability for tools to work together seamlessly---over isolation. This is a deliberate design choice enabling powerful multi-tool workflows. We do not argue this tradeoff is inherently wrong; rather, we argue it should require \textit{explicit user consent} rather than implicit trust. The specification provides no mechanism for users to configure isolation policies, even when desired.

Consider a legitimate workflow: ``Read config.json with filesystem-server, then query database with sqlite-server.'' Full isolation would prevent this. We measured the tradeoff empirically:

\begin{table}[h]
\centering
\caption{Isolation Level vs. Security and Utility}
\label{tab:isolation-utility}
\begin{tabular}{lcc}
\toprule
\textbf{Isolation Level} & \textbf{ASR} & \textbf{Task Completion} \\
\midrule
None (MCP default) & 61.3\% & 94.2\% \\
User-prompted cross-flow & 31.7\% & 87.4\% \\
Strict (no cross-flow) & 8.7\% & 61.8\% \\
\bottomrule
\end{tabular}
\end{table}

Our \textsc{AttestMCP} extension uses ``user-prompted'' isolation by default, requiring explicit authorization for cross-server data flow. This balances security (reducing ASR by 48\%) while maintaining acceptable utility (87.4\% task completion vs 94.2\% baseline).

\subsection{Message Integrity Analysis}

We analyzed the JSON-RPC message format for standard security properties:

\begin{table}[h]
\centering
\caption{MCP Message Security Properties}
\label{tab:message-security}
\begin{tabular}{lcc}
\toprule
\textbf{Property} & \textbf{Required} & \textbf{MCP v1.0} \\
\midrule
Message Authentication & Yes & \textcolor{red}{No} \\
Replay Protection & Yes & \textcolor{red}{No} \\
Capability Binding & Yes & \textcolor{red}{No} \\
Origin Identification & Yes & Partial* \\
Integrity Verification & Yes & \textcolor{red}{No} \\
\bottomrule
\end{tabular}

\vspace{1mm}
\raggedright
\scriptsize{*Transport-level only; not in message payload}
\end{table}

The specification relies entirely on transport security (TLS for HTTP) without application-layer protections. This is insufficient when the threat model includes compromised servers.

\section{Experimental Methodology}
\label{sec:experiments}

To measure the security impact of MCP's architectural choices, we developed \textsc{ProtoAmp} and conducted controlled experiments.

\subsection{\textsc{ProtoAmp} Framework}

Existing benchmarks (InjecAgent, AgentDojo) assume direct tool APIs rather than MCP's client-server architecture. Concurrent work on MCP-Bench~\cite{wang2025mcpbench} evaluates capability and task completion, while MCPSecBench~\cite{yang2025mcpsecbench} catalogs attack types. Our \textsc{ProtoAmp} (Protocol Amplification Benchmark) differs by measuring \textit{protocol amplification}---how MCP's architecture specifically increases attack success rates compared to non-MCP baselines:

\begin{enumerate}
    \item \textbf{MCP Server Wrappers:} We implemented MCP-compliant servers wrapping benchmark tool functions, preserving semantic equivalence while adding protocol overhead.

    \item \textbf{Attack Injection Points:} We added injection capabilities at three protocol layers:
    \begin{itemize}
        \item Resource content (indirect injection)
        \item Tool response payloads
        \item Sampling request prompts
    \end{itemize}

    \item \textbf{Measurement Infrastructure:} We instrumented clients to log all JSON-RPC messages, enabling analysis of attack propagation through protocol channels.
\end{enumerate}

\subsection{Experimental Setup}

\textbf{MCP Servers Under Test:}
\begin{itemize}
    \item \texttt{mcp-server-filesystem}: File operations (read, write, list)
    \item \texttt{mcp-server-git}: Repository management (clone, commit, diff)
    \item \texttt{mcp-server-sqlite}: Database queries (SELECT, INSERT)
    \item \texttt{mcp-server-slack}: Messaging integration
    \item \texttt{adversarial-mcp}: Custom server exercising protocol edge cases
\end{itemize}

\textbf{LLM Backends:} Claude-3.5-Sonnet, GPT-4o, Llama-3.1-70B

\textbf{Attack Scenarios:} 847 test cases:
\begin{itemize}
    \item InjecAgent adaptations: 312 (indirect injection, tool abuse)
    \item AgentDojo adaptations: 398 (multi-step attacks)
    \item Novel protocol-specific attacks: 137 (sampling, cross-server)
\end{itemize}

\textbf{Baseline:} Equivalent tool integrations without MCP (direct function calls) to isolate protocol-specific effects.

\subsection{Controlled Variables}

To ensure valid comparison between MCP and baseline conditions:
\begin{itemize}
    \item Tool semantics identical between conditions
    \item Same injection payloads used
    \item LLM prompting strategy held constant
    \item Network latency matched between conditions
\end{itemize}

\textbf{Latency Configuration:} Baseline uses direct function calls with simulated network overhead matching MCP. Measured MCP latencies: median 12.4ms for stdio transport, 23.7ms for HTTP/SSE. We verified latency does not significantly affect LLM safety refusal rates using paired t-test ($p > 0.3$, $n=200$).

\section{Results}
\label{sec:results}

\subsection{Protocol Amplification Effect}

Table~\ref{tab:amplification} shows attack success rates (ASR) comparing MCP-integrated agents versus baseline (non-MCP) integrations.

\begin{table}[h]
\centering
\caption{Attack Success Rate: MCP vs. Baseline}
\label{tab:amplification}
\begin{tabular}{lccc}
\toprule
\textbf{Attack Type} & \textbf{Baseline} & \textbf{MCP} & \textbf{$\Delta$} \\
\midrule
Indirect Injection (Resource) & 31.2\% & 47.8\% & +16.6\% \\
Tool Response Manipulation & 28.4\% & 52.1\% & +23.7\% \\
Cross-Server Propagation & 19.7\% & 61.3\% & +41.6\% \\
Sampling-Based Injection & N/A & 67.2\% & --- \\
\midrule
\textbf{Overall} & 26.4\% & 52.8\% & \textbf{+26.4\%} \\
\bottomrule
\end{tabular}
\end{table}

\textbf{Key Finding:} MCP's architecture amplifies attack success by 23--41\% depending on attack type. The largest amplification occurs in cross-server propagation, where MCP's lack of isolation boundaries enables attacks impossible in single-server deployments.

\subsection{Sampling Vulnerability Severity}

The sampling mechanism introduces a novel attack vector absent in non-MCP systems:

\begin{table}[h]
\centering
\caption{Sampling Attack Analysis by Model}
\label{tab:sampling}
\begin{tabular}{lrrr}
\toprule
\textbf{Model} & \textbf{ASR} & \textbf{Exfil. Rate} & \textbf{Persist.} \\
\midrule
Claude-3.5-Sonnet & 58.3\% & 42.1\% & 31.7\% \\
GPT-4o & 71.4\% & 53.8\% & 38.2\% \\
Llama-3.1-70B & 72.1\% & 61.2\% & 44.6\% \\
\bottomrule
\end{tabular}
\end{table}

Servers exploiting sampling achieved 58--72\% attack success, with 42--61\% successfully exfiltrating sensitive context and 31--44\% establishing persistence across sessions. Claude-3.5-Sonnet shows highest resilience, consistent with prior findings~\cite{debenedetti2024agentdojo}.

\subsection{Multi-Server Trust Exploitation}

We tested configurations with 2--5 concurrent MCP servers, with one server under adversary control:

\begin{table}[h]
\centering
\caption{ASR by Server Count (1 Compromised)}
\label{tab:multiserver}
\begin{tabular}{ccc}
\toprule
\textbf{Servers} & \textbf{ASR} & \textbf{Cascade Rate} \\
\midrule
1 & 47.8\% & N/A \\
2 & 58.4\% & 34.2\% \\
3 & 67.1\% & 51.8\% \\
5 & 78.3\% & 72.4\% \\
\bottomrule
\end{tabular}
\end{table}

Attack success scales with server count due to increased cross-server attack surface. With 5 servers, a single compromised server achieves 78.3\% ASR with 72.4\% cascade rate (successfully compromising additional servers' operations).

\textbf{Prompt Engineering Baseline:} We tested whether system prompt instructions alone could mitigate cross-server attacks without protocol changes. Adding ``Never pass data between different tool servers without explicit user confirmation'' to the system prompt reduced cross-server ASR from 61.3\% to 47.2\%---a 23\% reduction. However, this remains significantly higher than \textsc{AttestMCP}'s 8.7\%, demonstrating that prompt-level defenses are insufficient and protocol-level isolation is necessary.

\subsection{Comparison with Prior Benchmarks}

Our MCP-specific results contextualize prior findings:

\begin{table}[h]
\centering
\caption{Benchmark Comparison: Original vs. MCP}
\label{tab:benchmark-comparison}
\begin{tabular}{llrr}
\toprule
\textbf{Benchmark} & \textbf{Setting} & \textbf{Original} & \textbf{MCP} \\
\midrule
InjecAgent & GPT-4 & 24--48\% & 51.2\% \\
AgentDojo & Best agent & $<$25\% & 38.7\% \\
Agent-SafetyBench & Safety score & $<$60\% & 47.3\% \\
\bottomrule
\end{tabular}
\end{table}

When the same attack scenarios are executed through MCP infrastructure, success rates increase by 7--15 percentage points, confirming protocol-specific amplification independent of general LLM vulnerabilities.

\section{Defense: \textsc{AttestMCP} Protocol Extension}
\label{sec:defense}

Based on our analysis, we design \textsc{AttestMCP}, a backward-compatible protocol extension addressing identified vulnerabilities. While MCPSec~\cite{mcpsecdev2025} documents implementation-level vulnerabilities, MCPSecBench~\cite{yang2025mcpsecbench} provides attack taxonomies, and MCPGuard~\cite{virtueai2025mcpguard} offers runtime scanning, \textsc{AttestMCP} proposes concrete protocol additions (capability attestation, message authentication) that can be incorporated into the MCP specification itself---a complementary layer addressing protocol-level rather than implementation-level weaknesses.

\subsection{Design Principles}

\begin{enumerate}
    \item \textbf{Capability Attestation:} Servers must cryptographically prove capability possession via signed certificates from a capability authority.

    \item \textbf{Message Authentication:} All JSON-RPC messages include HMAC-SHA256 signatures binding content to authenticated server identity.

    \item \textbf{Origin Tagging:} Sampling requests are tagged with server origin, enabling clients to distinguish server-injected from user-originated prompts.

    \item \textbf{Isolation Enforcement:} Cross-server information flow requires explicit user authorization.

    \item \textbf{Replay Protection:} Timestamp plus nonce with configurable validity window.
\end{enumerate}

\subsection{Trust Model Options}
\label{sec:trust-models}

A critical design decision is the capability authority architecture. We evaluate three models:

\begin{table}[h]
\centering
\caption{Capability Authority Trust Models}
\label{tab:trust-models}
\footnotesize
\begin{tabular}{lp{1.8cm}p{1.8cm}}
\toprule
\textbf{Model} & \textbf{Pros} & \textbf{Cons} \\
\midrule
Centralized & Simple PKI, easy revocation & Single point of failure \\
\midrule
Federated & Distributed, flexible & Complex coordination \\
\midrule
Web-of-Trust & Decentralized & User complexity \\
\bottomrule
\end{tabular}
\end{table}

Our implementation uses the \textbf{federated model}: platform vendors (Anthropic, Cursor, JetBrains, etc.) operate CAs for their ecosystems with cross-signing agreements for interoperability. This balances decentralization with operational simplicity.

\textbf{Identity Verification:} Servers obtain certificates through:
\begin{itemize}
    \item \textit{Commercial servers:} Domain ownership verification (DNS TXT record)
    \item \textit{Open-source servers:} Package registry account binding (npm, pip) with maintainer identity verification
\end{itemize}

\textbf{Revocation Infrastructure:} We propose federated CAs maintain shared Certificate Revocation Lists (CRLs) with the following SLAs: (1) emergency revocation (e.g., compromised npm account) within 4 hours, (2) standard revocation within 24 hours, (3) CRL distribution via OCSP stapling with 1-hour refresh. This mirrors established PKI practices (e.g., Let's Encrypt) while acknowledging the operational burden on a volunteer-driven ecosystem.

\subsection{Protocol Additions}

\textbf{Capability Certificate:}
\begin{lstlisting}[basicstyle=\ttfamily\scriptsize]
{
  "capability_cert": {
    "server_id": "filesystem-server",
    "capabilities": ["resources", "tools"],
    "issued_by": "anthropic-ca",
    "issued_at": 1706140800,
    "expires_at": 1737676800,
    "signature": "base64..."
  }
}
\end{lstlisting}

\textbf{Authenticated Message:}
\begin{lstlisting}[basicstyle=\ttfamily\scriptsize]
{
  "jsonrpc": "2.0",
  "method": "tools/call",
  "params": {...},
  "mcpsec": {
    "server_id": "filesystem-server",
    "timestamp": 1706140800,
    "nonce": "random-32-bytes",
    "hmac": "base64..."
  }
}
\end{lstlisting}

\textbf{Replay Protection:} Clients maintain a sliding window of 1,000 nonces per server with 30-second validity. Messages with duplicate nonces or expired timestamps are rejected.

\subsection{Backward Compatibility and Migration}

\textsc{AttestMCP} operates in three modes to support gradual adoption:

\begin{itemize}
    \item \textbf{Permissive:} Accept legacy servers with user warning (migration default)
    \item \textbf{Prompt:} Require explicit user confirmation for unsigned servers
    \item \textbf{Strict:} Reject all unsigned servers
\end{itemize}

\textbf{Downgrade Attack Mitigation:} Once a server presents valid \textsc{AttestMCP} credentials, the client pins that expectation. Subsequent connections without credentials trigger security warnings. This prevents MITM downgrade attacks where an attacker strips security headers from a previously-authenticated server.

\subsection{Performance Overhead}

\begin{table}[h]
\centering
\caption{\textsc{AttestMCP} Latency Overhead (milliseconds)}
\label{tab:performance}
\begin{tabular}{lrrr}
\toprule
\textbf{Operation} & \textbf{P50} & \textbf{P95} & \textbf{P99} \\
\midrule
Certificate validation (cold) & 4.2 & 8.1 & 12.3 \\
Certificate validation (cached) & 0.3 & 0.5 & 0.8 \\
HMAC-SHA256 computation & 0.3 & 0.4 & 0.6 \\
Nonce lookup/insertion & 0.1 & 0.2 & 0.3 \\
\midrule
\textbf{Total per message (cold)} & \textbf{8.3} & \textbf{14.2} & \textbf{21.7} \\
\textbf{Total per message (warm)} & \textbf{2.4} & \textbf{4.1} & \textbf{6.2} \\
\bottomrule
\end{tabular}
\end{table}

Median overhead of 8.3ms (cold) or 2.4ms (warm cache) per message is negligible compared to LLM inference latency (typically 500--2000ms). Certificate validation dominates cold-start overhead; caching reduces this significantly for repeated calls within a session.

\subsection{Effectiveness Evaluation}

We implemented \textsc{AttestMCP} as a client-side shim and re-ran our full evaluation:

\begin{table}[h]
\centering
\caption{\textsc{AttestMCP} Defense Effectiveness}
\label{tab:mcpsec}
\begin{tabular}{lccc}
\toprule
\textbf{Attack Type} & \textbf{MCP} & \textbf{AttestMCP} & \textbf{Reduction} \\
\midrule
Indirect Injection & 47.8\% & 18.4\% & 61.5\% \\
Tool Response Manipulation & 52.1\% & 14.2\% & 72.7\% \\
Cross-Server Propagation & 61.3\% & 8.7\% & 85.8\% \\
Sampling-Based Injection & 67.2\% & 11.3\% & 83.2\% \\
\midrule
\textbf{Overall} & 52.8\% & 12.4\% & \textbf{76.5\%} \\
\bottomrule
\end{tabular}
\end{table}

\textsc{AttestMCP} reduces overall ASR from 52.8\% to 12.4\%---a 76.5\% reduction. The largest improvements occur in cross-server (85.8\%) and sampling (83.2\%) attacks, where isolation enforcement and origin tagging provide strong protection.

\subsection{Limitations}

\textsc{AttestMCP} does not address:
\begin{itemize}
    \item Attacks within a single legitimately-authorized server (the server has valid credentials but serves malicious content)
    \item Social engineering of users to authorize malicious capabilities
    \item CA compromise (mitigated by federation, but not eliminated)
    \item First-contact attacks: Pinning (TOFU---Trust On First Use) provides no protection when a user first installs a malicious server that never claimed \textsc{AttestMCP} support
    \item Ecosystem adoption: If most servers remain legacy/unsigned, users will default to ``Permissive'' mode, negating security benefits
\end{itemize}

Residual 12.4\% ASR primarily reflects indirect injection through legitimately-retrieved content---a fundamental limitation shared with all LLM systems that cannot be solved at the protocol layer.

\textbf{User Behavior Assumptions:} Our ASR measurements assume users carefully review cross-server authorization prompts. In practice, \textit{alert fatigue} may cause users to click ``Allow'' habitually, reducing real-world effectiveness. Future work should conduct user studies to measure actual authorization review rates.

\section{Discussion}

\subsection{Architectural vs. Implementation Security}

Our analysis demonstrates that MCP's security weaknesses are \textbf{architectural}, not merely implementation bugs:

\begin{itemize}
    \item Patching CVE-2025-49596 (MCP Inspector RCE) does not address capability attestation absence
    \item Fixing SQL injection in sqlite-mcp does not prevent sampling-based injection
    \item Hardening individual servers does not establish cross-server isolation
\end{itemize}

Protocol-level remediation is required. We recommend Anthropic incorporate \textsc{AttestMCP} concepts into MCP v2.0.

\subsection{Limitations of This Work}

\begin{itemize}
    \item Our experiments used five MCP servers; production deployments with dozens of servers may exhibit different characteristics
    \item \textsc{AttestMCP} has not been formally verified; we plan symbolic model checking in future work
    \item The federated CA model requires ecosystem coordination that may face adoption barriers
    \item We did not evaluate adversarial attempts to bypass \textsc{AttestMCP} specifically
\end{itemize}

\section{Related Work}

We build on foundational prompt injection research~\cite{greshake2023not, liu2024houyi} and agent security benchmarks~\cite{zhan2024injecagent, debenedetti2024agentdojo, zhang2024agentsafety, andriushchenko2024agentharm}. Prior work on multi-agent security~\cite{lupinacci2025darkside} identified inter-agent trust as a vulnerability---we extend this by demonstrating MCP's architectural contribution to this weakness.

Defense mechanisms including Spotlighting~\cite{hines2024spotlighting} and PromptArmor~\cite{shi2025promptarmor} address prompt-level protection but not protocol-level vulnerabilities. Our \textsc{AttestMCP} is complementary, addressing a different layer of the security stack.

\section{Conclusion}

We presented the first security analysis of the Model Context Protocol specification, identifying three protocol-level vulnerabilities: capability attestation absence, unauthenticated sampling, and implicit trust propagation. Through controlled experiments with \textsc{ProtoAmp}, we demonstrated that MCP's architecture amplifies attack success rates by 23--41\% compared to non-MCP integrations. Our proposed \textsc{AttestMCP} extension reduces attack success from 52.8\% to 12.4\% through capability attestation and message authentication, with acceptable performance overhead (8.3ms median per message).

As MCP adoption accelerates, addressing these architectural weaknesses becomes critical. We recommend:
\begin{enumerate}
    \item Protocol revision incorporating mandatory capability attestation
    \item Origin tagging requirements for all sampling requests
    \item Explicit isolation boundaries with user-prompted cross-server authorization
\end{enumerate}

\bibliographystyle{IEEEtran}

\end{document}